\documentclass[sigconf]{acmart}

\usepackage{booktabs} 
\usepackage{graphicx}
\usepackage{amsfonts}
\usepackage{amsmath}
\usepackage{amssymb}
\usepackage[colorinlistoftodos]{todonotes}
\usepackage{pdfpages}
\usepackage{algorithm}
\usepackage{float}
\usepackage[noend]{algpseudocode}

\usepackage{xspace} 

\DeclareMathOperator*{\argmax}{argmax} 

\newcommand{\attackname}{MaskDGA\xspace}
\newcommand{\etal}{\emph{et al. }}

\newcommand{\rot}[1]{\rotatebox[origin=c]{90}{#1}}
\newcommand{\rowgroup}[1]{\hspace{-1em} \textbf{#1}}

\setcopyright{none}

\begin{document}
\title[\texorpdfstring{\attackname}{\space}: A Black-box Evasion Technique Against DGA Classifiers]{\texorpdfstring{\attackname}{\space}: A Black-box Evasion Technique Against DGA Classifiers and Adversarial Defenses}

\author{Lior Sidi}
\affiliation{%
  \institution{Ben-Gurion University of the Negev}
}
\email{liorsid@post.bgu.ac.il}

\author{Asaf Nadler}
\affiliation{%
  \institution{Ben-Gurion University of the Negev
  }
}
\email{asafnadl@post.bgu.ac.il}

\author{Asaf Shabtai}
\affiliation{%
  \institution{Ben-Gurion University of the Negev}
}
\email{shabtaia@bgu.ac.il}

\begin{abstract}
Domain generation algorithms (DGAs) are commonly used by botnets to generate domain names through which bots can establish a resilient communication channel with their command and control servers.
Recent publications presented deep learning, character-level classifiers that are able to detect algorithmically generated domain (AGD) names with high accuracy, and correspondingly, significantly reduce the effectiveness of DGAs for botnet communication. 
In this paper we present \attackname, a practical adversarial learning technique that adds perturbation to the character-level representation of algorithmically generated domain names in order to evade DGA classifiers, without the attacker having any knowledge about the DGA classifier's architecture and parameters.
\attackname was evaluated using the DMD-2018 dataset of AGD names and four recently published DGA classifiers, in which the average F1-score of the classifiers degrades from 0.977 to 0.495 when applying the evasion technique.
An additional evaluation was conducted using the same classifiers but with adversarial defenses implemented: adversarial re-training and distillation. 
The results of this evaluation show that \attackname can be used for improving the robustness of the character-level DGA classifiers against adversarial attacks, but that ideally DGA classifiers should incorporate additional features alongside character-level features that are demonstrated in this study to be vulnerable to adversarial attacks.

\end{abstract}



\maketitle

\section{Introduction} \label{sec:introduction}
Botnets are groups of inter-connected devices that are designed to carry large-scale cyber-attacks~\cite{wiki:botnets} such as distributed denial-of-service~\cite{alomari2012botnet}, data-theft~\cite{nadler2019detection}, and spam~\cite{pathak2009botnet}. 
The design of modern botnets often involve advanced techniques that challenge the ability to detect the botnet operation and take it down. 
Domain generation algorithms (DGAs)~\cite{Plohmann2016a}, a notorious technique, has been used by more than 40 documented botnets in the last decade.

Domain generation algorithms are used to generate a large number of pseudo-random domain names, which are usually based on the date and a secret input (seed).
A bot and its command and control server that wish to communicate will both execute the DGA with a shared seed in order to generate a sequence of domain names and identify the one through which their communication can take place. 
DGAs can be used to generate thousands of domain names per day which must be identified and analyzed in order to shutdown of the botnet.

The detection of algorithmically generated domain (AGD) names initially focused on capturing binary samples of bots, extracting their algorithms and seeds, and generating the domain names in advance for mitigation~\cite{porras2009conficker, stone2009your, plohmann2015dgaarchive}.
However, by using new input seeds, this approach can be easily evaded by botnets.
In fact, between the years of 2017 and 2018 at least 150 new seeds were introduced by botnets,\footnote{Based on the change-log of https://data.netlab.360.com/dga/} a figure which is more than two times the number of documented DGAs. 
Using machine learning in order to inspect the lexicographic patterns of domain names for detecting AGD names and classify their generating algorithms~\cite{Tran2018a,Woodbridge2016a} is an alternative and more generalized approach.

Machine learning-based DGA detection techniques have become extremely successful, with state of the art algorithms achieving high detection rates on multiple datasets~\cite{sivaguruevaluation} with inline latency~\cite{koh2018inline}.
These techniques reduce the effectiveness of DGAs for maintaining resilient and stealthy botnet communication. 
From the botnet operator's perspective, a solution that can evade these state of the art detection mechanisms by using adversarial machine learning to produce AGD names that are less likely to be detected can be beneficial.

Adversarial machine learning is a technique in the field of machine learning which attempts to ``fool'' models (during either the training or execution phases) through adversarial input~\cite{45816}, also referred to as adversarial samples. 

In the context of DGA classification, generative adversarial networks (GANs)~\cite{goodfellow2014generative} were previously suggested by Anderson \etal~\cite{Anderson2016} in a method called DeepDGA, in order to to generate adversarial samples of ``fake'' domain names that are indistinguishable from real benign domain names that were available in the training set.
The fake domain names generated are used to augment the training set with additional benign domain names and accordingly improve the model's robustness.
Because DeepDGA was trained to generate domain names that resemble benign domains, the fake domain names generated are mostly short and readable.
This makes the generated domain names impractical for botnets, since they are more likely to be used by legitimate users (e.g., the domain name ``laner.com'' generated by DeepDGA was already owned by Laner Electric Supply).
In addition, short domain names are usually more expensive and thus increase the attacker's costs for maintaining the botnet communication.

\begin{figure*}[h!]
  \centering
    \includegraphics[width=0.9\textwidth]{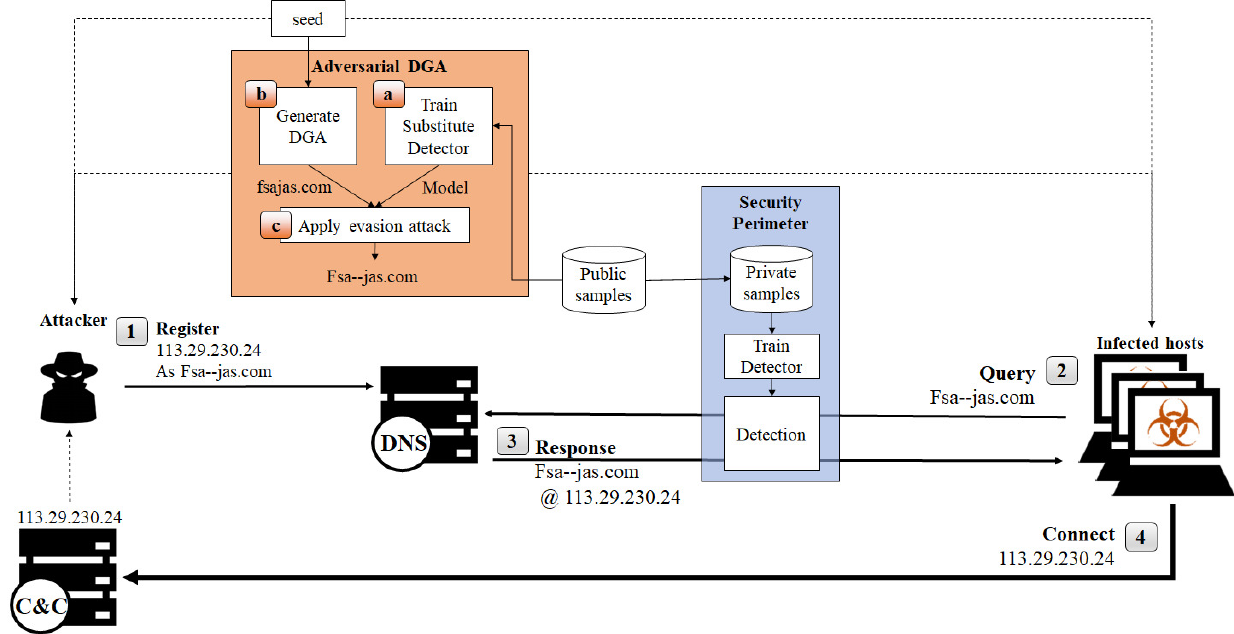}
    \caption{After training a substitute model (a), and generating a set of AGD names using an existing DGA (b), the attacker applies \attackname as a wrapper to add perturbation to the set of AGD names, which will result in a new set of adversarial domain names that evade detection (c).
    The bot operator can register the adversarial AGD names (1) to allow its bots to establish a communication channel with their command and control server via the adversarial domain names (2-4) while evading detection on the security perimeter.}
    \label{fig:hl_design}
\end{figure*}

In this paper we explore the ability to evade detection by DGA classifiers.
We present \attackname, a black-box adversarial learning technique that adds perturbations to a character level representation of an AGD and transforms it to a new domain name that is falsely classified as benign by DGA classifiers, without prior knowledge of the classifier's architecture and parameters.
\attackname can be applied as an extension to an existing malware to evade detection on the network perimeter by security solutions (as depicted in Figure~\ref{fig:hl_design}).

\attackname utilizes the transferability property~\cite{Papernot2016a} and is conducted by first training a substitute model on datasets of publicly known AGD names. 
The attacker then generates a set of AGD names to which \attackname is applied.
The domain names generated are provided as input to the substitute model and applied in a single feed forward step. 
The results of the feed forward step are used to compute the loss with regards to the benign class. 
\attackname then performs a single back-propagation step in which the loss propagates the gradients back to the input to form a Jacobian saliencey map~\cite{papernot2016limitations} (JSM).
Finally, for every domain name generated, \attackname replaces exactly half of the characters that had the largest gradient values in the JSM, so that the resulting domain name remains long and likely unreadable.
Changing only half of the characters for a set of randomly generated AGD names also implies that we can expect a small number of collisions among the domain names in the output set, and therefore, the number of domains in the output set will remain close to that of the original set (created by the DGA).
Eventually, the effort required for detecting and taking down the botnet remains the same.

\attackname was designed based on the Jacobian-based saliency map adversarial attack~\cite{papernot2016limitations} (JSMA) that was originally applied to images.
In contrast to the JSMA attack, \attackname is restricted to (1) producing valid character-level changes, (2) performing a constant number of changes, and (3) changing the same feature (character) only once, thus making it more suitable for character-level DGA classification evasion.

The evaluation of \attackname was performed using the recently published DMD-2018 dataset of AGD names and open-source DGA classifiers~\cite{ChoudharyAlgorithmicallyClassification}.
We also applied DeepDGA~\cite{Anderson2016} as a technique for evading the detection models, and compared its performance with that of \attackname.
The results show that when applying \attackname, the detection rate degrades from an average F1-score of 0.977 to 0.495, while the DeepDGA attack results in an average F1-score of 0.780.

We also evaluated \attackname against the same DGA classifiers enhanced with adversarial learning defenses, namely adversarial re-training (using domains generated by DeepDGA~\cite{anderson2016deepdga} and by \attackname) and distillation~\cite{Papernot2016DistillationNetworks}, and assessed their effectiveness.
The results shows that adversarial re-training can be effective to some extent against the evasion techniques but still leads to the conclusion that a more resilient detection mechanism is required.

In summary, the contributions of this paper are as follows:
\begin{itemize}
    \item we present a practical evasion technique for DGA detection machine learning models;
    \item the evasion technique can be applied as a wrapper on the existing DGA source code used by the botnet and therefore easy to apply;
    \item the technique is generic and can be applied to any DGA family and input seed, while adhering to the expected requirements of a DGA, i.e., uniqueness, validity and rareness of domain names;
    \item we present an evaluation of the presented evasion technique using four recently published DGA classifiers, and we assess the robustness of \attackname against commonly used adversarial defense techniques;
    \item from the presented evasion technique we conclude that the detection of DGA mechanisms should not rely solely on character-level DGA features and ideally DGA classifiers should incorporate additional behavioral features.
\end{itemize}

\section{Background}
\subsection{Character-level DGA classification} \label{subsec:character-level}
A character-level representation of a word $W$ over the alphabet of symbols $V$ is a binary matrix $X_{|W| \cdot |V|}$ s.t.
\[ \forall i \in [W], j \in [V] : X_{i,j} = \left\{
\begin{array}{ll}
      1 & W_i = V_j \\
      0 & \text{otherwise} \\
\end{array} 
\right. \]
i.e., $X_{i,j}$ is set to one if the $i$-th character of the word $W$ matches the $j$-th symbol in the alphabet $V$ and zero if otherwise.

Recently published models for DGA classification~\cite{Woodbridge2016a, Saxe2017} are based on character-level representations of domain names due to their high accuracy~\cite{Bielik2017} and inline latency~\cite{koh2018inline}, provided that no additional context is required (e.g., network traffic features). 
Therefore, the focus of the evasion techniques presented in this study is on character-level DGA classification.

\subsection{Adversarial learning}
Adversarial learning attacks can be divided into poisoning attacks performed during training~\cite{biggio2011bagging, biggio2012poisoning}, and evasion attacks performed during testing~\cite{biggio2013evasion, goodfellowexplaining, papernot2017practical}.  
Evasion attacks usually involve adding perturbation (noise) to the input sample's features to result in a misclassification by the attacked model.
Such inputs are also referred to as adversarial samples.
In order to make the targeted model classify the sample $X$ whose true label is $Y$ as $Y^{*}$, the attacker would have to find a perturbation $\delta_X$ such that $(X+\delta_X)$ is classified as $Y^{*}$.
If the attack is aimed at a specific class in the data, it is referred to as a targeted attack~\cite{Papernot2016a}.

Attacks also vary based on the knowledge the adversary has about the classifier. 
A black-box evasion attack requires no knowledge about the model beyond the ability to query it as a black-box, i.e., inserting an input and getting the output classification. 
In a white-box attack, the adversary can have varying degrees of knowledge about the model architecture, hyper parameters and data used for training. 

Most black-box attacks rely on the concept of adversarial example transferability, also known as the transferability property~\cite{Papernot2016a}.
According to the transferability property, adversarial examples crafted against one model are also likely to be effective against other models, even when the models are trained on different datasets (but assumed to be sampled from the same distribution)~\cite{papernot2016limitations,szegedy2013intriguing}.
This means that the adversary can train a substitute model, which has decision boundaries similar to the original model, and perform a white-box attack on it, i.e., obtain the back-propagated gradients in order to add perturbation that would result in the desired target misclassification.
Adversarial examples that successfully fool the substitute model are likely to fool the original model as well~\cite{papernot2017practical}. 
In this paper we present a black-box technique aiming at evading DGA detection models.

\section{Threat Model} \label{sec:attacker_settings}
\paragraph{Training capability (black-box)} We assume that the targeted DGA detection models are not available to the botnet operator, and therefore the attacker cannot rely on knowledge about the architecture of the targeted model. 
However, we assume that the botnet operator can use publicly available AGD datasets and previously published architectures in order to train a substitute model. 

\paragraph{Targeted misclassification for the benign class} The targeted models are multiclass DGA classification models whose input is a domain name and output is a probability distribution for the set of DGA classes and the benign class (e.g., ~\cite{Woodbridge2016a, Saxe2017}).
The class with the maximal value in the output probability distribution is selected by the DGA classifier as the predicted class.
A misclassification between the various DGA labels would still result in the detection of the malicious domain name and is therefore considered an unsuccessful adversarial attack. 
Thus, the attacker's goal is a targeted misclassification of the adversarial input as the benign class.

\section{The Evasion Technique} \label{sec:attack}
\subsection{Overview}
\begin{figure*}[h!]
  \centering
    \includegraphics[width=0.9\textwidth]{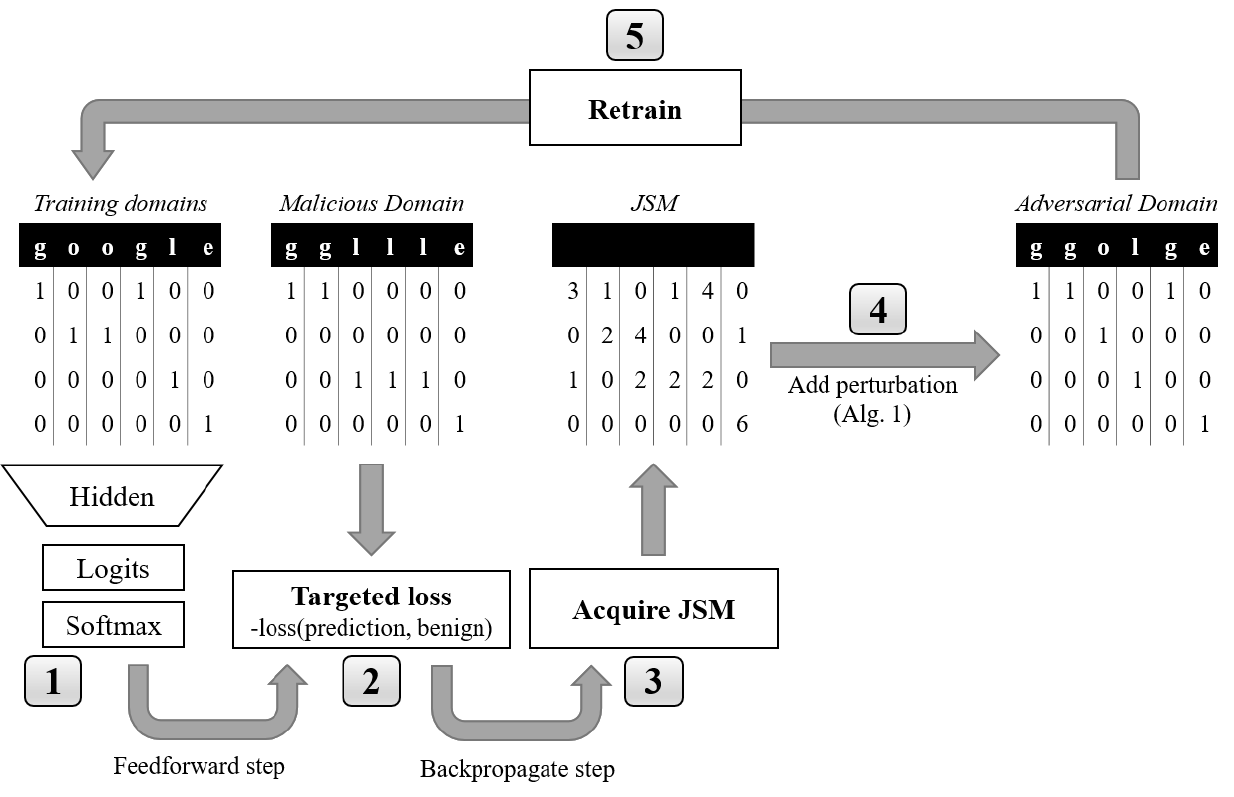}
    \caption{The attacker trains a substitute model on a publicly available dataset of AGD (1).
    A single feed forward step is performed using the substitute model on an input AGD name to compute the targeted loss with regard to the benign class (2).
    The targeted loss with regards to the benign class is used to back-propagate the gradients and construct the Jacobian-based saliency map (3) and add perturbation to the input AGD based on the JSM to result in a new adversarial domain names that evades detection (4).}
    
    \label{fig:method}
\end{figure*}

The evasion technique is performed according to the following main steps (portrayed in Figure~\ref{fig:method}):

\begin{itemize}
    
    \item \textbf{Training a substitute model.} The attacker acquires a 
    dataset of public samples of AGD names and uses it to train a discrminative model that distinguishes AGD names from benign domain name. 
    The model is referred to as the substitute model.
    
    \item \textbf{Generating the algorithmically generated domain names.} The attacker executes an existing DGA with a randomly selected seed and generates a set of AGD names.
    
    \item \textbf{Constructing a Jacobian-based saliency map.} For every generated AGD, the attacker performs a single feed forward step in the substitute model. 
    The results of the feed forward step are used to compute the loss with regard to the benign class. 
    The attacker performs a single back-propagation step on the loss in order to acquire the Jacobian-based saliency map, which is a matrix that assigns every feature in the input AGD with a gradient value.
    Features (characters) with higher gradient values in the JSM would have a more significant (salient) effect on the misclassification of the input AGD name and hence the name saliency map.
    
    \item \textbf{Adding perturbation to the input AGD name based on the JSM.} For every input AGD name, iterate over every character position from the first character to the last. 
    For every position, save the highest salient symbol and its value, e.g., in the first position the symbol ``a'' may have the highest value in the JSM. 
    Set the top half of the character positions with regard to the maximal saliency values to the symbols that yielded the maximal saliency values (as portrayed in Algorithm~\ref{alg:attack}).
    
\end{itemize}

\subsection{Training a substitute model} \label{subsec:substitute_model}

The substitute model is a discrminative model (i.e., classifier) that distinguishes algorithmically generated domain names from benign domain names. 
The input layer of the model accepts character-level domain name features in the feed forward step and then reads the gradient for every feature in the back-propagation step for the construction of the JSM.
The output layer of the model can be either binary (e.g., DGA or benign) or multiclass to predict the specific DGA. 
Formally, the substitute model learns a function $F$ that accepts an input domain name $X$ with a predicted class $Y$. 
These notations are later used in Algorithm~\ref{alg:attack}.

\subsection{Generating the domain names} \label{subsec:generating_domains}

The generation of original AGD names can be executed either by botnets that already have a DGA mechanism implementation and are looking to extend it to evade detection, or by a new DGA implementation.
Although the evasion technique works regardless of the selected DGA, the attacker should take into account two considerations: the average length and the number of generated domain names.

The average length of the generated domain names represents a trade-off between the ease of evasion and the availability of the domain name generated for botnet communication. 
AGD names that are shorter (usually less than eight symbols) have a higher chance of being readable.
Readable domain names have greater likelihood of evading detection, but on the other hand, they are more likely to be owned by legitimate users and thus unavailable for attacks. 
In addition, \attackname might make the provided domain names even more readable (while maintaining their length) and accordingly less available. 
In contrast, longer domain names are likely to be available and inexpensive.
However, DGA classifiers are sensitive to the length of the domain names and are more likely to predict longer than average domain names as AGD names, regardless of their characters.

The number of generated domain names represents another trade-off between stealthy communication and resilient communication. 
DGAs that produce a very large number of domain names (e.g., thousands per day) are more difficult for take-down but are easier to identify because of the overall noise.
An alternative is a DGA that produces a smaller number of domain names to choose a more stealthy communication.
For instance: Banjori, Conficker and Gameover Zeus DGAs which generate thousands of domain names per day are ideal in terms of resilience, while Torpig and Matsnu generate up to three domain names per day and are better for stealthy communication~\cite{Plohmann2016a}.

\subsection{Construct a Jacobian-based saliency map} \label{subsec:jsm}

The construction of the JSM begins by providing the substitute model with an AGD name $X$ as performing a single feed forward step.
For every AGD name, the feed forward step would provide a predicted probability distribution for the output classes of the substitute model.
Based on the predicted probability distribution and the benign class distribution, we compute the \emph{targeted} loss function $\mathcal{L}(F(X), Y^{*})$ of the predicted value $F(X)$ with respect to the benign class $Y^{*}$.
The back-propagation algorithm is then applied on the computed loss from the last hidden layer to the input layer so that every feature of the input $X$ is assigned with a gradient.
For example, the feature $X_{i,j}$ will be assigned with the real-value gradient $\frac{\partial \mathcal{L}(F(X), Y^{*})} {\partial X_{0,0}}$.

Formally, the JSM (denoted as $S$) is the set of the back-propagated gradients for every input feature:
\[
S = 
\begin{bmatrix}
   \frac{\partial \mathcal{L}(F(X), Y^{*})} {\partial X_{0,0}} & \frac{\partial \mathcal{L}(F(X), Y^{*})} {\partial X_{1,0}} & \dots & \frac{\partial \mathcal{L}(F(X), Y^{*})} {\partial X_{|W|,0}} \\
\vdots & \vdots & \ddots & \vdots \\
\frac{\partial \mathcal{L}(F(X), Y^{*})} {\partial X_{0,|V|}} & \frac{\partial \mathcal{L}(F(X), Y^{*})} {\partial X_{1,|V|}} & \dots & \frac{\partial \mathcal{L}(F(X), Y^{*})} {\partial X_{|W|,|V|}} \\
\end{bmatrix}
\]

\subsection{Adding perturbation to the input AGD based on the JSM} \label{subsec:translate}

The input character-level AGD $X$ is added with a perturbation $\delta_X$ so that following constraints are met:
\begin{enumerate}
    \item $X + \delta_X$ is a valid character-level representation of a domain name (see Subsection~\ref{subsec:character-level})
    \item Exactly half of the characters of $X$ are replaced  
\end{enumerate}

The selection the perturbation $\delta_X$ and the addition of its to transform $X$ to $X'$ is formally described in Algorithm~\ref{alg:attack}. 
Initially, the algorithm computes the maximal value for every column in the JSM. 
The median of maximal value acts as a threshold and is henceforth denoted as $\Theta$.
\attackname replaces the character in position $i$ to the $j$-th symbol in the alphabet if and only if the maximal value for column $i$ of the JSM is $j$ and $j > \Theta$. 
The character replacement is performed by adding a perturbation $\delta_X$ in which:
\begin{enumerate}
    \item the original symbol is set to $(-1)$,
    \item the new symbol is set to $(+1)$, and
    \item the rest of the symbols are set to zero.
\end{enumerate}

By definition, since $\Theta$ is the median of the maximal columns in the JSM that correspond to character positions, exactly half of the characters are replaced. 
The result is a new domain name $X^{*}$ which is the original $X$ with the perturbation $\delta_X$ that was induced by the algorithm.

\begin{algorithm}
\caption{Adding perturbation to an algorithmically generated domain name based on the JSM}
\label{alg:attack}
\begin{flushleft}
        \textbf{INPUT:} \\
        $X$ - a character-level AGD name of size $|W| \cdot |V|$ \\
        $Y^{*}$ - the targeted class \\
        $F$ - is the function learned by the substitution model
\end{flushleft}
\begin{algorithmic}[1]
\State Let $S$ be the saliency map of $\nabla F(X)$ w.r.t. $Y^{*}$
\State $X^{*} \gets X$
\For{$i$ in $0,..,|W|$} 
    \State $S^{i}_{j^*} \gets \max_{j} S_{i,j}$ 
\EndFor
\State Let $\Theta$ be the median of $\{ S^{0}_{j^*}, ..., S^{|W|}_{j^*} \}$ \Comment{Set the threshold}

\For{$i$ in $0,..,|W|$}
    \If{$\argmax_{j} S_{i,j} \ge \Theta$}
    \State $X^{*}_{i,j} = 1$ 
    \State $X^{*}_{i,\overline{j}} = 0$ for all $\overline{j} \neq j$
    \EndIf
\EndFor

\State \textbf{return} $X^{*}$
\end{algorithmic}
\end{algorithm}

\section{Evaluation}
\subsection{Overview}
In this section we present the experiments conducted in order to evaluate the \attackname evasion technique.
\attackname was tested on four recently presented models which are based on deep learning DGA classifiers using a public dataset of AGD names.\\
The experiments focused on evaluating the effectiveness of \attackname while targeting the attacked models in general, as well as when focusing on a specific DGA family. \\
The performance of \attackname was compared against two baseline attacks: (1) a random attack, which selects half of a domain's characters at random and replaces them with uniformly selected characters from the alphabet of English letters, digits, hyphens and underscores, and (2) an attack based on DeepDGA~\cite{Anderson2016}. \\
The DeepDGA-based attack was conducted using the source code provided by the authors of~\cite{Anderson2016} with several modifications that were required in order to use it for the attack.
After upgrading the source code so we could use the newer version of the Keras library~\cite{chollet2015keras} and applying it to a recent Alexa top 500,000 sites (as was done by Anderson \etal~\cite{Anderson2016}), the generator produced a limited variety of samples.
This is a common challenge in applying GANs, also known as \emph{mode collapse}, which expresses the sensitivity of GAN architectures to minor changes in the configuration such as when using a different dataset~\cite{hui2018gans}.
Our understanding of the problem was that the generator was quickly over-fitted to the discriminator and failed to generate new samples.
To overcome the mode collapse state we provided the discriminator with an initial advantage over the generator by pre-training the discriminator on a preliminary set of domain names that were generated by the generator before the training phase.

\subsection{Datasets} \label{subsec:dataset}
The main dataset used for the evaluation is the DMD-2018 dataset~\cite{Vinayakumar2018BigApplications,vinayakumar2018scalable,vinayakumar2018detecting,mohan2018spoof} which contains 100,000 benign domain names and 297,777 algorithmically generated domains that were evenly produced by twenty different DGA families (see Table~\ref{tbl:dataset}). 
The substitute model that used in \attackname was trained on 113,335 randomly selected domain names (28.9\% of the dataset).
The targeted models (i.e., the four DGA classifiers) that we attempted to evade were trained on 264,456 domain names (67.6\% of the dataset) to provide the DGA classifiers with an advantage against the attack.
The remaining 9,986 domain names (3.3\% of the dataset) were used to test the effectiveness of \attackname. 
Every domain name in the test set was perturbed and evaluated against the targeted models.

Additionally, the alexa.com dataset of top sites~\footnote{https://www.alexa.com/topsites} was used to train the initial autoencoder of DeepDGA~\cite{anderson2016deepdga} as conducted by Anderson \etal~\cite{Anderson2016}.
Using the trained DeepDGA, we generated 500 adversarial domain names that were used to compare with \attackname, and an additional 63,500 domain names were used for the a defense as explained in Subsection~\ref{sec:defenses}.

\begin{table}[ht!]
\centering
\caption{The dataset used for evaluating \attackname.} \label{tbl:dataset} 
\begin{tabular}{llll}
\toprule
Class       & Substitute training & Targeted training & Testing  \\
\midrule
\rowgroup{Benign} \\
benign      & 27000            & 63000          & - \\
\midrule
\rowgroup{DGA} \\
banjori     & 4349             & 10148          & 503   \\
corebot     & 4349             & 10148          & 503   \\
dircrypt    & 4349             & 10148          & 503   \\
dnschanger  & 4349             & 10148          & 503   \\
fobber      & 4349             & 10148          & 503   \\
murofet     & 4349             & 10148          & 503   \\
necurs      & 3704             & 8644           & 429   \\
newgoz      & 4349             & 10148          & 503   \\
padcrypt    & 4349             & 10148          & 503   \\
proslikefan & 4349             & 10148          & 503   \\
pykspa      & 4349             & 10148          & 503   \\
qadars      & 4349             & 10148          & 503   \\
qakbot      & 4349             & 10148          & 503   \\
ramdo       & 4349             & 10148          & 503   \\
ranybus     & 4349             & 10148          & 503   \\
simda       & 4349             & 10148          & 503   \\
suppobox    & 4349             & 10148          & 503   \\
symmi       & 4349             & 10148          & 503   \\
tempedreve  & 4349             & 10148          & 503   \\
tinba       & 4349             & 10148          & 503  \\
\midrule
Total DGA      & 86335           & 201456       & 9986 \\
Total       & 113335           & 264456       & 9986 \\
\bottomrule
\end{tabular}\par
\bigskip
The number of domain names in the dataset that are used for the evaluation. 
The substitute training set is used by the attacker to train the substitute model. 
The targeted training set is used to train a DGA classifier model that is targeted by the attacker. 
The test set contains an additional set of adversarial AGD names that are tested against the DGA classifiers.
\end{table}

\subsection{Substitute model architecture}
The architecture of the substitute model has the character-level (one hot encoding) representation of the domain as the input layer and an output neuron for every DGA family or benign domain as the output layer.
The hidden layers consist of one-dimensional CNN filters of size 2, 3, 4, 5, and a max pooling layer that reduces the domain length to one half of its original and two additional fully-connected layers.
This architecture was chosen because we wanted to evaluate the effectiveness of \attackname when using a model that is considerably weaker then the attacked DGA detection classifiers, and because it is lightweight and thus practical to apply.

The training of the substitute model was conducted for five epochs with batches of 128 domain names until convergence using the ADAM optimizer~\cite{kingma2014adam} with a learning rate of 0.01. 

The implementation of the evaluated adversarial evasion technique was achieved using the Keras library~\cite{chollet2015keras} and the TensorFlow library~\cite{abadi2016tensorflow} for the substitute model and the Cleverhans library~\cite{papernot2018cleverhans} for computing the gradients based on the targeted loss, and accordingly the JSM.

The character-level representations inputs for domain names are based on the vocabulary of English letters, digits, hyphens, underscores and dots.
We (hard-coded) restrict the \attackname from adding perturbations that would change a non-dot character to a dot character.
This restriction is designed to avoid the \emph{shortening} of the second-level domain (e.g., ``example.com'' turns into ``exa.mple.com'') which has undesired effects on the effectiveness of the evasion technique (as explained in Section~\ref{sec:introduction}).

\subsection{Targeted (attacked) models' architecture}

The targeted models that \attackname tries to evade from detection are based on the following recently proposed DGA classifiers:

\begin{itemize}
    \item Endgame~\cite{Anderson2016} which is based on a single LSTM layer (denoted as LSTM[Endgame])
    \item CMU~\cite{dhingra2016tweet2vec} which is based on a forward LSTM layer and a backward LSTM layer (denoted as biLSTM[CMU])
    \item Invincea~\cite{Saxe2017} which based on parallel CNN layers (denoted as CNN[Invincea])
    \item MIT~\cite{vosoughi2016tweet2vec} which based on stacked CNN layers and a single LSTM layer (denoted as CNN + LSTM[MIT])
    
\end{itemize}

The targeted models were implemented using the TensorFlow library~\cite{abadi2016tensorflow} and are based on the source code that was published by~\cite{Bielik2017} and trained on the the set of domain names that are reserved for the targeted model training in the dataset.

\subsection{Evaluation metrics}
The metrics used to evaluate the effectiveness of the evasion techniqued on the targeted models are precision, recall, and the F1-score, as is commonly used for DGA classification evaluation.
Based on the threat model (see Section~\ref{sec:attacker_settings}) where the goal is defined as targeted misclassification, we define the correct class as any DGA, instead of the specific DGA i.e., the classifier might predict the wrong DGA class but it is regarded correct as long as it did not misclassify an AGD as benign.
\begin{description}
	\item[Precision] computes the model's ability to assign the correct label out of all the instances in cluster label.
	\begin{equation}
	Precision(C,Y) = \frac{correct\, class}{predict\, class}
	\label{eq:prec}
	\end{equation} 
	
	\item[Recall] computes the model's ability to retrieve all of the instances of a certain class; i.e., the model's ability to retrieve all of the domains of a certain family. 
	\begin{equation}
	Recall(C,Y) = \frac{correct\, class}{entire\, class}
	\label{eq:rec}
	\end{equation}
	
	\item[F1 score] combines precision and recall together into one score using harmonic mean 
	\begin{equation}
	F1(C,Y) = 2 * \frac{precision(C,Y) * recall(C,Y)}{precision(C,Y) + recall(C,Y)}
	\label{eq:f1}
	\end{equation} 
\end{description}

\subsection{Results} \label{subsec:attackresults}

Table~\ref{tbl:eval_per_attack} presents the precision, recall and F1 measures when applying the evasion techniques against the four tested models (LSTM[Endgame], biLSTM[CMU], CNN[Invincea], and CNN + LSTM[MIT]).
We compare the performance of the targeted models when no attack is applied ($No Attack$), when a random attack is performed ($Random$), when a DeepDGA-based attack is performed ($DeepDGA$), and when executing the presented evasion technique ($\attackname$).

Based on the results, we can see that when no adversarial attack is applied ($No Attack$), all of the models perform almost perfectly on the test set.
More specifically, the targeted models yield an average precision (0.977), recall (0.977) and F1-score (0.977), thus supporting the claim that recently published, deep learning based DGA classifiers are very accurate in detecting AGD names.

The $Random$ attack shows a degradation in the average precision (-19\%), recall (-26\%), and F1-score (-26.5\%), compared to the $No Attack$ case.
This degradation in the evaluation measures indicates that the four targeted models are very sensitive to changes in the lexical patterns of DGA-based domain names.

The DeepDGA attack consists of generating pseudo-random domain names that are sampled from the same distribution of the benign Alexa top 1M domains using a GAN network. 
While clearly degrading the performance of the detection models in comparison to the $No Attack$ case, the DeepDGA generated domains are effectively detected by all four models with a higher average precision (+6\%), recall (+5\%) and F1-score (+0.3\%) compared to the $Random$ attack.
The generation of domain names by DeepDGA relies on sampling symbols from a pre-trained multinomial regressor that provides the likelihood for every symbol at a given position.
The sampling is performed independently for every position, and thus it is not guaranteed that the output domain name would not appear to be algorithmically generated.
The \attackname evasion technique dominates all evaluated attacks on all four targeted models with an average recall rate of 0.575; i.e., an average DGA classifier would miss 42.5\% of the AGD names compared to only 2.5\% in the $No Attack$ case.

\begin{table}[ht!]
\centering
\caption{Evaluation results.} \label{tbl:eval_per_attack} 
\begin{tabular}{llll}
\toprule
 Targeted model & Precision & Recall & F1-score \\
  \midrule 
  \rowgroup{CNN[Invincea]} \\
No Attack  & 1.00              & 1.00           & 1.00              \\
Random     & 0.83             & 0.74          & 0.73            \\
DeepDGA    & 0.91             & 0.80           & 0.84            \\
\attackname    & \textbf{0.68}             & \textbf{0.52}          & \textbf{0.39}                  \\
\midrule
\rowgroup{LSTM[Endgame]} \\
No Attack  & 0.98             & 0.98          & 0.98                  \\
Random     & 0.83             & 0.79          & 0.78                 \\
DeepDGA    & 0.74             & 0.73          & 0.74                 \\
\attackname    & \textbf{0.73}             & \textbf{0.60}           & \textbf{0.53}      \\
\midrule
\rowgroup{biLSTM[CMU]} \\
No Attack  & 0.98             & 0.98          & 0.98\\
Random     & 0.85             & 0.81          & 0.81            \\
DeepDGA    & 0.75             & 0.82          & 0.78            \\
\attackname    & \textbf{0.72}             & \textbf{0.61}          & \textbf{0.55} \\
\midrule
\rowgroup{CNN + LSTM [MIT]} \\
No Attack  & 0.95             & 0.95          & 0.95      \\
Random     & 0.80              & 0.77          & 0.77     \\
DeepDGA    & 0.68             & 0.92          & 0.74     \\
\attackname    & \textbf{0.63}             & \textbf{0.57}          & \textbf{0.51} \\
\bottomrule
\end{tabular}\par
\bigskip

The evaluation metrics of the targeted DGA classifiers on the test set when no attack is applied ($No Attack$), a random attack is performed ($Random$), a DeepDGA-based attack is performed ($DeepDGA$), and when executing the presented evasion technique ($\attackname$). 
\end{table}

We were also interested in evaluating the effectiveness of the evasion technique on specific DGA families that are characterized by different patterns. \\
Table~\ref{tbl:eval_per_dga} presents the F1-score (which reflects the harmonic mean between the precision and recall) for the twenty DGA families that are available in the test set.

When $No Attack$ is applied, a similar average F1-score for all DGA families is observed.
This indicates that the models memorize the lexicographical patterns of observed domains regardless of their length.

The $Random$ attack, when focusing on the specific DGA families, results in a more significant degradation for some DGAs (e.g., Simda, Suppobox) than others (e.g., CoreBot, Symmi). 
The Suppobox DGA is a dictionary-based DGA and therefore its generated domains appear as normal English terms (e.g., tablethirteen.net).
Therefore, a random replacement of half of the characters would completely demolish the lexicographical pattern of Suppobox domains and accordingly the $Random$ attack is highly effective for Suppobox. 
However, for DGAs such as CoreBot which produces long domains with uniformly selected characters (e.g., i8a0q2wdu8otulkfylo2gdq.ddns.net) the $Random$ attack is not effective against all models (e.g., LSTM[Invincea] results with an average F1-score of 0.91). \\
The presented \attackname evasion technique is dominant over all DGA families across all targeted models. 
When \attackname is applied to the Suppobox dictionary-based DGA all targeted models detect the adversarial domain name with an average F1-score of 0.495.
For the more difficult case of the corebot DGA, the average F1-score is 0.7725, thus arguing that while the longer AGD have more difficulty in evading detection, \attackname still manages to evade detection. \\
Note that in this analysis the DeepDGA could not be evaluated (and thus was omitted) because its input is a seed string and not AGD that were produced by DGA processes. 

\begin{table*}[ht!]
\footnotesize
\centering
\caption{Evaluation results per DGA family (average F1-score).} \label{tbl:eval_per_dga} 
\begin{tabular}{lllllllllllllllllllll}
\toprule
Model/DGA   & \rot{banjori} & \rot{corebot} & \rot{dircrypt} & \rot{dnschanger} & \rot{fobber} & \rot{murofet} & \rot{necurs} & \rot{newgoz} & \rot{padcrypt} & \rot{proslikefan} & \rot{pykspa} & \rot{qadars} & \rot{qakbot} & \rot{ramdo} & \rot{ranbyus} & \rot{simda} & \rot{suppobox} & \rot{symmi} & \rot{tempedreve} & \rot{tinba} \\ \midrule
    \rowgroup{CNN[Invincea]} \\
No Attack                         & 0.97                           & 0.97                           & 0.97                            & 0.97                              & 0.97                          & 0.97                           & 0.96                          & 0.97                          & 0.97                            & 0.97                               & 0.97                          & 0.97                          & 0.97                          & 0.97                         & 0.97                           & 0.97                         & 0.97                            & 0.97                         & 0.97                              & 0.97                         \\
Random                            & 0.78                           & 0.91                           & 0.74                            & 0.79                              & 0.79                          & 0.89                           & 0.78                          & 0.88                          & 0.72                            & 0.81                               & 0.80                           & 0.77                          & 0.83                          & 0.78                         & 0.73                           & 0.7                          & 0.70                             & 0.92                         & 0.77                              & 0.76                         \\
\attackname                           & 0.49                           & 0.67                           & 0.50                             & 0.52                              & 0.49                          & 0.64                           & 0.50                           & 0.56                          & 0.49                            & 0.51                               & 0.51                          & 0.55                          & 0.54                          & 0.54                         & 0.49                           & 0.51                         & 0.49                            & 0.49                         & 0.51                              & 0.53 \\ \midrule
\rowgroup{LSTM[Endgame]} \\
No Attack                         & 0.90                            & 0.90                            & 0.89                            & 0.89                              & 0.90                          & 0.90                            & 0.88                          & 0.88                          & 0.90                             & 0.89                               & 0.89                          & 0.9                           & 0.9                           & 0.9                          & 0.9                            & 0.89                         & 0.89                            & 0.9                          & 0.9                               & 0.9                          \\
Random                            & 0.78                           & 0.89                           & 0.69                            & 0.63                              & 0.84                          & 0.84                           & 0.76                          & 0.85                          & 0.71                            & 0.75                               & 0.73                          & 0.78                          & 0.81                          & 0.84                         & 0.83                           & 0.65                         & 0.69                            & 0.88                         & 0.75                              & 0.74                         \\
\attackname                           & 0.48                           & 0.82                           & 0.55                            & 0.51                              & 0.67                          & 0.68                           & 0.6                           & 0.65                          & 0.51                            & 0.56                               & 0.58                          & 0.64                          & 0.64                          & 0.63                         & 0.64                           & 0.52                         & 0.51                            & 0.68                         & 0.63                              & 0.59   \\     \midrule
\rowgroup{LSTM[CMU]} \\
No Attack                         & 0.87                           & 0.87                           & 0.87                            & 0.86                              & 0.87                          & 0.87                           & 0.85                          & 0.87                          & 0.87                            & 0.86                               & 0.87                          & 0.87                          & 0.87                          & 0.87                         & 0.87                           & 0.87                         & 0.87                            & 0.87                         & 0.87                              & 0.87                         \\
Random                            & 0.8                            & 0.87                           & 0.78                            & 0.73                              & 0.81                          & 0.81                           & 0.66                          & 0.85                          & 0.71                            & 0.69                               & 0.74                          & 0.77                          & 0.81                          & 0.84                         & 0.71                           & 0.76                         & 0.7                             & 0.87                         & 0.78                              & 0.67                         \\
\attackname                           & 0.49                           & 0.86                           & 0.61                            & 0.58                              & 0.64                          & 0.66                           & 0.53                          & 0.67                          & 0.49                            & 0.53                               & 0.57                          & 0.58                          & 0.63                          & 0.6                          & 0.55                           & 0.57                         & 0.5                             & 0.81                         & 0.62                              & 0.54        \\
\midrule 
\rowgroup{CNN[MIT]} \\
No Attack                         & 0.76                           & 0.76                           & 0.75                            & 0.76                              & 0.76                          & 0.76                           & 0.74                          & 0.75                          & 0.76                            & 0.75                               & 0.75                          & 0.76                          & 0.76                          & 0.76                         & 0.76                           & 0.75                         & 0.76                            & 0.76                         & 0.76                              & 0.76                         \\
Random                            & 0.66                           & 0.76                           & 0.63                            & 0.59                              & 0.71                          & 0.72                           & 0.64                          & 0.74                          & 0.63                            & 0.65                               & 0.63                          & 0.69                          & 0.69                          & 0.71                         & 0.67                           & 0.59                         & 0.64                            & 0.75                         & 0.65                              & 0.64                         \\
\attackname                           & 0.48                           & 0.74                           & 0.5                             & 0.49                              & 0.56                          & 0.62                           & 0.53                          & 0.63                          & 0.48                            & 0.54                               & 0.51                          & 0.52                          & 0.56                          & 0.51                         & 0.53                           & 0.49                         & 0.48                            & 0.64                         & 0.52                              & 0.51  \\
\bottomrule
\end{tabular} \par
\bigskip
The average F1-score of the targeted DGA classifiers on twenty DGA families when: no attack is applied ($No Attack$), a random attack is performed ($Random$), and when executing the presented evasion technique ($\attackname$). 
\end{table*}

Finally, Table~\ref{tbl:examples} contains examples of AGD that were generated by several DGA families and their adversarial form after adding perturbation by \attackname in order to provide further intuition.
As described in Section~\ref{sec:attack}, the characters that are replaced by \attackname are the ones that have the highest saliency with regards to a misclassification as the benign class. 
Accordingly, the provided examples inflict two interesting trends.
The clearest trend is that the underscore (``\_'') characters is frequently used by \attackname since it is valid and common for benign domain names that appeared on the DMD-2018 dataset but \emph{never} appeared on any in the dataset, and at all to the best of our knowledge. \\
Another trend is the use of digits, especially in the prefix of domain names which is relatively rare for AGD. 
Also, note that the pykspa domain ``abfmfid.net'' resulted in the adversarial domain name ``poqhfid.net'' that involves a consonant-vowel-consonant (CVC) pattern that is common in English readable words, and accordingly in benign domains.

\begin{table}[hb!]
\centering
\caption{Examples of adversarial domain names generated by \attackname} \label{tbl:examples} 
\begin{tabular}{@{}ccc@{}}
\toprule
DGA           & Original                 & After \attackname             \\ \midrule
qadars      & 02sygu4egq8m.net         & \_2\_yq-apgq8m.net          \\
fobber      & coclocpxjwgoefjih.net    & \_oclocpxjwfqjf\_fh.net     \\
proslikefan & csdwxmk.biz              & lqjw1mk.biz                 \\
pykspa      & abfmfid.net              & poqhfid.net                 \\
tinba       & bcpprwxhxktb.pw          & 2\_\_\_rncqxktb.pw          \\ \bottomrule
\end{tabular}\par
\bigskip
Example of algorithmically generated domains from six DGA families in their original and adversarial form.
\end{table}

\section{Evaluation of Adversarial Defenses} \label{sec:defenses}
\subsection{Overview}
The emergence of adversarial attacks was followed by studies that suggested appropriate defenses. 
In this section we evaluate the effectiveness of two common adversarial defenses, namely adversarial re-training and distillation, against the presented \attackname evasion technique.

\subsection{Tested defenses}
The first defense evaluated is adversarial re-training.
In adversarial re-training~\cite{YuanAdversarialLearning} the training set is augmented with adversarial samples so that the resulting model is more robust to adversarial samples.
While re-training is considered an effective defense against adversarial samples, it requires prior knowledge of the defender about the attack and it does not necessarily generalize to other attacks~\cite{goodfellowexplaining}. \\
We tested two different adversarial re-training processes: re-training using \attackname (denoted as \textit{\attackname re-train}) and re-training using DeepDGA (denoted as \textit{DeepDGA re-train}).
\vspace{5mm}

In \textbf{\attackname re-train} each DGA classifier (i.e., the targeted model) is initially trained on the same training set as described in Section~\ref{subsec:dataset}. 
Then, we randomly selected one hundred AGD names from each of the twenty DGA families (2,000 domains in total) and applied the \attackname on them.
The targeted model was then re-trained for two additional epochs on a new training set that consists of the original training set and the 2,000 adversarial samples. 
\vspace{5mm}

In order to apply \textbf{DeepDGA re-train}, we first trained the DeepDGA GAN architecture using 500k benign domain names and output 63,500 pseudo-random domain names. 
As suggested by Anderson \etal~\cite{Anderson2016}, we extend the training set of the targeted model with these domain names which are labeled as the ``benign'' class, in order to defend against adversarial attacks.
\vspace{5mm}

\textbf{Distillation networks}~\cite{Papernot2016DistillationNetworks} were proposed as a defense against adversarial perturbation that uses a special parameter called the ``distillation temperature'' $T$. 
Setting high temperature values (e.g., $T=10$): for the softmax layer as induces smoother decision bounds as follows:
\[
\text{Softmax}(X, T) = \bigg[ \frac{e^{z_i(X)/T}} {\sum^{N-1}_{l=0} e^{z_l(X)/T}} \bigg]_{i \in 0..N-1}
\]
where Softmax$(X, T)$ is the softmax layer, $T$ is the temperature, and $z_i$ is the $i$-th logit. 
After training the initial network with a high distillation temperature, the softmax values obtained (denoted as ``soft labels'') are now used to train a new network (denoted as the ``distilled network'') which is more resistant to attacks.

As a general adversarial defense, distillation networks have been evaluated on adversarial attacks against the MNIST~\cite{lecun2010mnist} and CIFAR-10~\cite{krizhevsky2014cifar} datasets and for both the effectiveness of the attacks dropped from a 95\% of misclassification  rate to less than 5\%. 
The main intuition regarding the reduced effectiveness of attacks is that the distilled soft labels allow samples that are in between labels (e.g., a figure of the digit '3' that resembles an '8') to be classified near the decision boundary instead of in the center, which makes the trained model vulnerable to attacks.

Our implementation of the distillation defense relies on the implementation of~\cite{carlini2017towards}, in which the same temperature was used for both the initial and distilled networks, respectively named teacher and student. 
Moreover, setting the temperature to values that are larger than 40, as was done on images in~\cite{Papernot2016DistillationNetworks}, failed to converge for DGA detection in which the classes of samples more often overlap. 
Therefore, we set the temperature to 10. 

\subsection{Results}
Table~\ref{tbl:eval_per_defense} presents the average F1-score measure of the four models in the cases of $No Attack$, $DeepDGA$ attack, and $\attackname$ attack, and when no defense is applied ($No Defense$), when using $\attackname$ $retrain$, $DeepDGA$ $retrain$, and $Distillation$. 

\begin{table}[hb!]
\centering
\caption{Evaluation of the effectiveness of the adversarial defenses (average F1-score).} \label{tbl:eval_per_defense} 
\begin{tabular}{llll}
\toprule
Defense / Attack              & No Attack & DeepDGA & \attackname \\
\midrule
\rowgroup{CNN[Invincea]} \\
No defense         & 1.00        & 0.84           & 0.39        \\
Distillation            & 1.00        & 0.85           & 0.42        \\
DeepDGA retrain       & 0.87       & \textbf{0.97}           & 0.40        \\
\attackname re-train & 0.96       & 0.72           & \textbf{0.96}        \\
\midrule
\rowgroup{LSTM[Endgame]} \\
No defense         & 0.98       & 0.74           & 0.53        \\
Distillation            & 0.98       & 0.74           & 0.52        \\
DeepDGA retrain       & 0.96       & \textbf{0.88}  & 0.57        \\
\attackname retrain & 0.95       & 0.77           & \textbf{0.91}        \\
\midrule
\rowgroup{biLSTM[CMU]} \\
No defense         & 0.98        & 0.78           & 0.55        \\
Distillation            & 0.98        & \textbf{0.86}  & 0.51        \\
DeepDGA retrain       & 0.94        & 0.48           & 0.54        \\
\attackname retrain & 0.93        & 0.64           & \textbf{0.85}        \\
\midrule
\rowgroup{CNN + LSTM[MIT]} \\
No defense         & 0.95        & 0.74           & 0.51        \\
Distillation            & 0.58        & 0.45           & 0.58        \\
DeepDGA retrain       & 0.91        & \textbf{0.92} & 0.49        \\
\attackname retrain & 0.91        & 0.54           & \textbf{0.80}       
\end{tabular} \par
\bigskip
Evaluation of the average F1-score of targeted models when applying the distillation, DeepDGA retrain and adversarial retrain defenses. 
\end{table}

The results presented in Section~\ref{subsec:attackresults} indicate that, when no defense is applied, the models are vulnerable to domain names that were generated by the DeepDGA attack and even more vulnerable to to domain names that were generated by \attackname.

The distillation network defense slightly improves the F1-scores against all attacks and across all models but is inefficient overall.
The primary reason for the inefficiency of the distillation defense is that it is designed for general misclassification (i.e., misclassification for any other class other than the correct class $Y$) and not a targeted misclassification (i.e., a misclassification for a specific class $Y'$).

The DeepDGA retrain defense significantly improve the robustness of the targeted models against the DeepDGA attack for three of the four models (all but CMU).
The DeepDGA hardening is ineffective against the \attackname evasion technique, but it comes with the cost of reducing the effectiveness of detecting non-adversarial AGD names.

The \attackname retrain defense provides the highest average F1-score results on all of the attacks and for all targeted models.
The effectiveness of detecting non-adversarial AGD names is still reduced in comparison to the targeted models without any defense, but overall it has better results on average, than the alternatives of DeepDGA hardening and distillation networks.
The effectiveness of the defense against the DeepDGA attack has an average F1-score of 0.6675, which is better than the effectiveness of the DeepDGA retrain defense against the \attackname evasion technique.
This suggests that overall, the defenses do not fully prevent general adversarial attacks against DGA classifiers.

\section{Practical implementation of \attackname}

Resilient communication between bots and their command and control server using a DGA is normally established as follows. 
The bot operator and the bots have an implementation of the DGA mechanism in their source code and share a secret input seed. 
Periodically, the botnet operator executes the DGA with the secret input seed to generate a large number of domain names and attempts to register them. 
The domain names that are successfully registered, are configured by the bot operator to resolve to the command and control system. 
In turn, the bot executes the DGA with the shared secret input seed to generate the same list of domain names that was generated by the bot operator.
The bot iterates the generated domain names and attempts to connect each one of them until it receives a reply from from the command and control server.
The bot that executes the DGA might reside on hosts which have limited computational and storage resources (e.g., IoT devices or point-of-sale machines).
Therefore, it is important to understand the practical implementation of the presented evasion technique in order to understand its requirements and limitations.

\attackname requires the bot to generate a set of AGD, use a substitute model to perform a feed-forward step and back-propagation to acquire the JSM, and add perturbation based on the JSM.
We assume that bots that have previously used DGAs are capable of generating a set of AGD names. 
Therefore, the main focus of this section the requirements of representing the AGD names as character-level binary matrices, loading a previously trained substitute model, and computing the JSM. 
The list of requirements is summarized in Table~\ref{tbl:practical_requirements}.

The longest AGD name in the rich DMD-2018 dataset has 65 characters, and the number of allowed symbols for domain names include case-insensitive English letters, digits, hyphens, underscores and dots, for an overall of 39 symbols. 
There a binary character-level representation of $65 \cdot 39 = 2535$ entries of bits is sufficient for the representation every single AGD name.

The substitute model that we trained can be serialized to the HDF5 binary serialization format~\footnote{https://www.h5py.org/} and later re-loaded with TensorFlow. 
Since TensorFlow's development libraries often exceeds 100 MB, TensorFlow Lite~\footnote{https://www.tensorflow.org/lite/overview} can be used.
TensorFlow Lite weighs less than 300 KB and is designed to load pre-trained models on Android, embedded devices and Desktops and supports various programming languages such as:: C, C++, Java and Python.
The substitute model that was trained in Section~\ref{subsec:substitute_model} had 160,853 parameters and is serialized to a 10MB HDF5 file.

For every AGD name the attacker is required to perform a single feed-forward step and a single back-propagation which is negligible in computation resources (since no training is required). 
The result of the backpropagation is the JSM matrix of gradients that has a similar number of entries to that of the input representation (2,535), but in contrast, every entry has 16 bits instead of a single bit.
Therefore, the overall JSM representation has slightly less than 38.4KB per AGD name.

\begin{table}[]
\centering
\caption{Practical implementation requirements.} \label{tbl:practical_requirements} 
\begin{tabular}{|c|c|c|}
\hline
Requirement & Memory & Supported platforms \\ 
\hline

TF Lite & 300KB & Android / Embedded / Desktop \\
\hline

Model & 10MB & - \\
\hline

Input Repr. & 2.4KB / AGD & - \\ 
\hline

JSM Repr. & 38.4KB / AGD & - \\ 
\hline
\end{tabular} \par
\bigskip
The platform and memory requirements of a bot to execute \attackname.
A bot would load a pre-trained substitute model using TensorFlow Lite that is supported on various environments and represent every AGD input and output JSM representation as a binary character-level matrix in memory.
\end{table}

\section{Conclusions}

This study portrays an adversarial machine learning evasion technique on DGA classifiers. 
The evaluation of the evasion technique on four state of the art, recently published DGA classifiers indicates that their performance degrades significantly.
\attackname is evaluated along with two other attacks, DeepDGA and random, which also resulted in a degradation in the performance of the DGA classifiers, although to a lesser extent than that caused by the \attackname. 
The performance degradation caused by the random attack helps us argue that character-level DGA classifiers memorize specific lexicographic patterns of known DGA and thus are extremely vulnerable to adversarial attacks.

The GAN architecture used by DeepDGA (1) requires extensive hyper-parameter tuning, (2) does not easily converge, and therefore is unstable and unexpected, (3) requires an approximation technique to address the discrete values of the input (i.e., the domain name), and (4) was trained to generate domains based on legitimate character distribution, and therefore is likely to generate benign (already registered) domain names.
On the other hand, the \attackname technique is (1) designed specifically for generating malicious domains (i.e., the substitute model is trained using both legitimate and malicious domains) and is therefore more successful, (2) requires minimal tuning (it only requires training a simple substitute model), and (3) is applied as a wrapper on the original DGA mechanism used by the botnet.
Moreover, \attackname can be executed by bots with limited resources as long as they install TensorFlow lite (supported on multiple platforms) and load a pre-trained model that weighs less than 10MB, thus making the evasion technique practical for botnets.

Additionally, we tested \attackname and the DeepDGA attack against three adversarial defenses, namely: \attackname re-training, DeepDGA re-training and distillation networks.
We conclude that distillation is ineffective against targeted misclassifications. 
The DeepDGA retraining was effective against the DeepDGA attack but caused the targeted model to degrade against non-adversarial AGD names. 
Finally, \attackname re-training on the \attackname adversarial samples had better results on average, but it still does not provide a perfect defense against other adversarial attacks such as DeepDGA.
These results demonstrate the vulnerability of character-level DGA classifiers to adversarial attacks and potential solutions to that are further discussed in Section~\ref{sec:discussion}.

\section{Related works}
The concept of using adversarial learning for generating malicious adversarial samples was previously investigated by~\cite{Hu2017, Rosenberg2018, rosenberg2018low, Suciu2018} for generating malware that evade machine learning-based anti-malware software and~\cite{Gao2018,Dasgupta2018} for evading text classifiers. 

DeepDGA~\cite{Anderson2016} extended these studies to the area of DGA detection by proposing a generative adversarial learning application that generates pseudo-random domain names that appear legitimate and thus can be used to augment the benign class in the training set for classification robustness.
DeepDGA can be regarded as both an evasion technique that produces pseudo-random domain names and as a defense mechanism for classification robustness.
As an evasion technique, DeepDGA is impractical for botnets as the generated domains are short and readable and therefore either expensive or not available for carrying out large attacks by botnets.
Moreover, the performance of \attackname is superior to that of DeepDGA and simpler to implement and tune, since it is based on a discriminative model which is less complicated computationally than generative models.

Character-level adversarial attacks were previously suggested in HotFlip~\cite{Ebrahimi2014} to cause misclassification of words by text classifiers in the white-box adversarial setting by replacing, inserting or deleting a single character. 
In contrast, the presented evasion technique is applied in a black-box adversarial setting in which a substitute model is trained. 
Also, deleting or inserting characters is forbidden to avoid the generation of longer domain names which are sensitive to detection and short domain names which are more expensive and less available. 
In addition, in \attackname the replacement of characters is performed on half of the characters, where every character is replaced once, in order to preserve a unique set of generated domains. 
In contrast, HotFlip always focuses on the character that induces the highest gradient since there's no restriction of replacing multiple words to the same words. 
For example, the words ``open,'' ``pea,'' and ``ten'' can all change to the word ``pen'' in a single flip, but \attackname ensures that the chances of outputting the same domain is small, resulting in a larger unique domain names, and accordingly making it difficult to detect all of the generated domain names.

\section{Discussion and future work} ~\label{sec:discussion}
Character-level DGA classifiers gained popularity largely due to their simplicity, the availability of training data, and the ability to use them for real-time detection.
We therefore aimed at evaluating the robustness of these classifiers against evasion techniques.
Since the evaluation of \attackname indicates the vulnerability of these classifiers, we propose potential solutions for improved robustness.
Firstly, DGA classifiers can improve their robustness against adversarial attacks using additional contextual features (e.g., DNS traffic features and WHOIS features~\cite{curtin2018detecting, schiavoni2014phoenix, antonakakis2012throw}) are also being used for classification. 
Another approach for robustness should focus on DGA detection systems that would detect domain names with novel lexicographic representation that implies the existence of either a new DGA process, a new secret seed or an adversarial attack.

\attackname can be generalized for other security scenarios, as well as non-security scenarios, in which character-level classifiers attempt to predict the malice of a sample. 
For example, phishing URLs classifiers can re-crafted using the presented technique to evade detection, and malware strings can be automatically replaced to overcome signature-based anti-malware tools.
We believe that using \attackname for adversarial re-training can improve the robustness of the models in these cases as well.


\section{Acknowledgements}
The authors would like to thank Hyrum Anderson for answering our questions regarding the DeepDGA paper and disclosing the source code of DeepDGA.
Additionally, the authors would like to thank R Vinayakumar for sharing the DMD-2018 dataset that was used in this study.

\bibliographystyle{ACM-Reference-Format}
\bibliography{sample-sigconf}

\end{document}